# Generative AI in Computer Science Education: Accelerating Python Learning with ChatGPT

Ian McCulloh
Johns Hopkins University
3400 N Charles St
Baltimore, MD 21218
lmccull4@jhu.edu

Pedro Rodriguez
Johns Hopkins University
3400 N Charles St
Baltimore, MD 21218
prodrig4@jh.edu

Srivaths Kumar
Great Learning
India
srivaths.kumar@greatlearning.in

Manu Gupta
Great Learning
manu.g@mygreatlearning.com

Viplove Raj Sharma
Great Learning
viplove@mygreatlearning.com

Benjamin Johnson
Johns Hopkins University
bjohn159@jhu.edu

Anthony N. Johnson
Johns Hopkins University
ajohn260@jhu.edu

## ABSTRACT

The increasing demand for digital literacy and artificial intelligence (AI) fluency in the workforce has highlighted the need for scalable, efficient programming instruction. This study evaluates the effectiveness of integrating generative AI, specifically OpenAI's ChatGPT, into a self-paced Python programming module embedded within a sixteen-week professional training course on applied generative AI. A total of 86 adult learners with varying levels of programming experience completed asynchronous Python instruction in Weeks three and four, using ChatGPT to generate, interpret, and debug code. Python proficiency and general coding knowledge was assessed across 30 different assessments during the first 13 weeks of the course through timed, code-based evaluations. A mixed-design ANOVA revealed that learners without prior programming experience scored significantly lower than their peers on early assessments ($\eta^2$ = .10–.17). However, following the completion of the accelerated Python instruction module, these group differences were no longer statistically significant ($\eta^2 < .05$), indicating that the intervention effectively closed initial performance gaps and supported proficiency gains across all learner groups. These findings suggest that generative AI can support accelerated learning outcomes and reduce entry barriers for learners with no prior coding background. While ChatGPT effectively facilitated foundational skill acquisition, the study also highlights the importance of balancing AI assistance with opportunities for independent problem-solving. The results support the potential of AI-augmented instruction as a scalable model for reskilling in the digital economy.

## CCS Concepts

• **Computing methodologies** → **Artificial intelligence** → **Natural language processing** → **Language models**   • **Applied computing** → **Education** → **Interactive learning environments**. **Social and professional topics** → **Computing education** → **Computing education programs.**

## Keywords

Generative AI; ChatGPT; Python; Programming education; Self-paced learning; Adult learners.

## 1. INTRODUCTION

The rapid advancement of artificial intelligence (AI) technologies and the growing demand for digital skills across industries have intensified the global need for reskilling and upskilling the workforce. Among the most critical competencies in this transformation is programming literacy, particularly in widely used languages such as Python. Python's versatility and readability have made it the language of choice not only for software development but also for data science, automation, and AI applications. Independent market research contracted by Johns Hopkins University reveals that 35% of software related job postings require Python proficiency and the year-on-year projected growth in jobs requiring Python is +24%. Yet, traditional approaches to teaching programming—centered on lectures, rote memorization, and repetitive coding practice—often require extended instructional timelines and produce uneven learning outcomes, particularly for adult learners or professionals transitioning into technical roles.

To address these challenges, this study explores an alternative instructional model for teaching introductory Python programming, compressing a traditional fourteen-week semester into a one-week (ten-hour) intensive, asynchronous, self-paced online course. Rather than relying on conventional lecture-based instruction, the course leverages OpenAI's ChatGPT to support accelerated, inquiry-based learning. Students are guided to interact with ChatGPT to generate initial code snippets, interpret code functionality, and receive natural language explanations for syntax errors, logic bugs, and debugging strategies. This paradigm shift encourages learners to develop not only coding skills, but also essential metacognitive skills such as problem formulation, solution synthesis, and adaptive troubleshooting.

This approach aligns with broader industry trends. Many technology companies now encourage or mandate the use of generative AI tools like ChatGPT, Gemini and Copilot to boost software development productivity, reduce time-to-solution, and streamline error handling. These tools are rapidly becoming integral to modern software engineering workflows, signaling a change in how programming is practiced in real-world settings. As such, computer science education must evolve to prepare learners to effectively engage with these tools—not simply as passive recipients of information, but as active, critical participants in AI-augmented development processes.

Accelerating programming instruction using AI tools is particularly valuable in the context of national and international efforts to build a resilient, AI-literate workforce. Government, industry, and academic institutions alike recognize the urgency of developing scalable, efficient training models that can quickly onboard individuals into technical fields. By integrating generative AI tools into the learning experience, this instructional model provides a scalable pathway for reskilling adult learners, supporting diversity in the technical workforce, and ultimately contributing to economic competitiveness in the digital age.

In the sections that follow, we outline the course design, learning objectives, implementation strategy, and initial observations from deploying this model. We also discuss the pedagogical implications of relying on AI tools for core programming instruction and propose future directions for evaluating long-term learning outcomes.

## 2. BACKGROUND

Learning to program is a cognitively demanding endeavor that presents a variety of challenges across cognitive, syntactic, and affective domains. These challenges, well-documented in the literature, often hinder students' progress in acquiring foundational programming skills. Recent developments in generative artificial intelligence (AI), particularly large language models (LLMs) like ChatGPT, have introduced new opportunities to augment programming education. This section reviews the known barriers to learning computer programming and synthesizes emerging research on the use of generative AI to support and accelerate Python instruction.

A key difficulty in learning to program is the mastery of syntax and semantics. Novice programmers frequently make syntactic errors or misunderstand semantic rules, impeding their ability to construct meaningful code [1,2]. Rosminah and Zamzuri [2] highlighted that even after receiving instruction, students continue to face challenges in debugging syntactical mistakes, suggesting that traditional feedback methods may be insufficient. Debugging and error handling further compound these difficulties. Identifying and correcting errors in code is particularly challenging for novices, who may lack experience to systematically troubleshoot issues [1].

Understanding computational concepts such as variables, loops, and functions also presents barriers [3, 4, 5]. Chan Mow [3] found that students have persistent difficulty in grasping the scope and utility of functions, which are essential for modular programming. In some cases, students perceive programming as a form of "magic," relying on memorization rather than comprehension. Pears et al. [5] argue that this superficial understanding results in poor transfer of skills to new problem contexts. This issue is often exacerbated by anxiety and fear of failure. According to [2] and [3], students frequently lose motivation after encountering repeated failures, leading to avoidance behavior and reduced engagement with course materials.

Further complicating instruction is the lack of timely and actionable feedback. Programming requires immediate feedback to reinforce correct practices and identify misconceptions. Delayed feedback, can lead to frustration and stagnation in learning [6].

Perhaps the biggest difficulty in learning to program is the requirement for abstract thinking and logical reasoning. Students are expected to conceptualize intangible constructs and apply logical operators to solve problems—skills that are not inherently intuitive. Caspersen and Bennedsen [7] found that the ability to abstract is a strong predictor of programming success, particularly in object-oriented paradigms. A number of studies document challenges with students' ability to understand abstract programming concepts [3]. Without this foundational skill, students often struggle to progress beyond surface-level understanding [1, 3, 8]. Algorithmic thinking—devising structured, step-by-step solutions to problems—is another cornerstone of programming education. However, students often struggle with decomposing problems and formulating appropriate algorithms. Qian and Lehman [6] observed that novice programmers have difficulty understanding the underlying logic of problems, leading to ineffective or incomplete solutions.

Traditional teaching modalities such as using books, lectures, presentations, and demonstrations appear to be ineffective modalities for knowledge and skill acquisition [3, 9, 10]. Students also report challenges with the pacing of traditional courses, especially when exposed to a steep learning curve. Bennedsen and Caspersen [9] emphasized that rapid introduction of complex topics can overwhelm learners, underscoring the need for flexible, adaptive learning environments. Additionally, the absence of clear real-world applications can diminish student motivation [1], while the overwhelming variety of tools and languages can lead to confusion and decision paralysis [5]. Foundational mathematics skills are also correlated with programming success; students with stronger algebra backgrounds tend to perform better in introductory programming courses [9]. Cheah [3] speculates that traditional methods are not personalized to the learner and do not provide sufficient live interaction and dynamic elements to explain programming concepts. He further suggests that it would be ideal for an instructor "to give immediate feedback, and detailed explanation whenever is needed by the students, however, this is impossible due to time limitations, manpower, and the number of students available in a single class.

Considering these challenges, generative AI models such as ChatGPT offer new pathways for supporting programming education. ChatGPT can generate code, provide explanations, and deliver personalized, immediate feedback—capabilities that may directly address some of the difficulties. For example, Rajbhoj et al. [11, 12] found that ChatGPT can automate various software development tasks, suggesting its potential utility in demonstrating real-world programming applications to students. Phung et al. [13] conducted a systematic evaluation of ChatGPT (based on GPT-3.5) and GPT-4, comparing their performance with human tutors across various programming education scenarios. The study revealed that GPT-4 substantially outperforms its predecessor and approaches the effectiveness of human tutors in several contexts. This underscores the potential of advanced generative AI models to provide high-quality educational support. Nonetheless, the research also identifies areas where GPT-4 struggles, emphasizing the need for further refinement and the importance of human oversight in educational applications.

Several tools have been developed to integrate ChatGPT into programming environments. Chen et al. [14] introduced GPTutor, a Visual Studio Code extension that provides real-time code explanations. Their initial evaluations suggest that such tools can enhance learning by providing contextual, on-demand assistance. However, the use of ChatGPT is not without limitations, indicating a need for more extensive user studies to assess long-term educational benefits. Dunder et al. [15] found that while ChatGPT could solve simple programming problems, it struggled with more complex ones, indicating that it should be viewed as a supplementary aid rather than a replacement for traditional instruction. Furthermore, using ChatGPT as a teachable agent—as demonstrated by Chen et al. [16]—may support knowledge acquisition by encouraging students to explain programming concepts to the model. This learning-by-teaching approach was shown to improve students' ability to write logical and readable code, although it had a limited impact on their debugging skills, which are often acquired through repeated trial and error.

Chen et al. [16] explored the concept of using ChatGPT as a teachable agent in a learning-by-teaching paradigm within programming education. Their findings indicate that engaging with ChatGPT in this manner enhances students' knowledge acquisition and programming abilities, particularly in writing readable and logically sound code. However, the study also notes a limited impact on developing error-correction skills, as ChatGPT's tendency to generate correct code reduces opportunities for students to practice debugging. This suggests that while ChatGPT can support certain aspects of learning, it may inadvertently neglect critical skills such as debugging, which are essential for programming proficiency.

Collectively, these findings suggest that ChatGPT has the potential to enhance Python instruction by accelerating the feedback loop, personalizing assistance, and contextualizing abstract concepts. However, to ensure comprehensive learning, instructional design must address the limitations of AI tools, such as the risk of student overreliance and reduced practice with debugging or algorithmic formulation. As such, the integration of ChatGPT into asynchronous, self-paced Python courses should be approached with pedagogical intention, ensuring that it complements rather than replaces foundational teaching strategies.

## 3. METHODS

A total of 86 learners participated in a sixteen-week course on applied generative artificial intelligence (AI), offered through a university-affiliated professional training program. The course was designed to upskill a broad audience in advanced generative AI capabilities, including the development of software solutions involving large language models, LangChain and LangGraph workflows, agentic AI systems, and advanced retrieval-augmented generation (RAG) techniques. Notably, prior Python programming experience was not required for enrollment.

To ensure all participants had the foundational coding skills necessary for the remainder of the course, students completed asynchronous, self-paced Python instruction during Weeks three and four. This module focused on helping learners become proficient in basic programming using Python, with an emphasis on leveraging OpenAI's ChatGPT to support code generation, interpretation, and debugging. Through guided use of generative AI, students learned how to generate initial code snippets, understand their functionality, and handle errors, thereby enabling them to engage meaningfully with the advanced content in subsequent weeks.

Based on a pre-course survey assessing prior programming experience, learners were categorized into three groups:

• Programming Experience in Python ($n = 50$): Learners with prior Python experience, ranging from beginner to advanced levels. Self-reported years of experience were recorded.

• Programming Experience in Other Languages Only ($n = 22$): Learners with experience in languages such as Java, R, or C++, but no prior exposure to Python.

• No Programming Experience ($n = 14$): Learners who self-identified as having no background in any programming language.

All participants completed a series of 30 assessments throughout the course. The first three assessments evaluated students' abilities to develop conditional statements, looping statements, and functions. The subsequent four assessments evaluated students' abilities to evaluate code that included Python fundamentals, variables/types, functions, and control flow statements. Assessments eight and nine were practice quizzes that evaluated introductory concepts involving AI and generative AI respectively. At this point, students had completed their generative AI assisted training in Python and were deemed to be "proficient" to the level of an introductory undergraduate course in Python programming. The subsequent 21 assessments ranged in topics that included foundational AI concepts, natural language processing (NLP), the use of transformers in large language models (LLM), prompt engineering, LangChain workflows (classification, text generation, and summarization), exploitations and guardrails to secure LLMs, agentic AI using LangGraph, developing retrieval augmented generative (RAG) search applications. In addition to these assessments, teaching assistants qualitatively evaluated student mastery of programming concepts during two major course projects, which required practical application of skills in generative AI contexts. Each assessment included a combination of multiple-choice questions and code-based problem-solving tasks.

The course was delivered entirely online using a combination of pre-recorded video lectures (3.5 hours), seven short coding assignments (5 hours), and a Python proficiency exam used to assess proficiency for graduate pre-requisite requirements (1 hour). All materials, including assessments, were administered digitally via a learning management system. Learners were encouraged to complete assignments independently, although optional group discussion forums were available.

Participants were included in the analysis if they had at least one valid assessment score. For repeated measures analysis, pairwise deletion was used: only those comparisons for which both timepoints were available for a learner were retained. Scores of 0 were interpreted conservatively as valid responses unless identified as clearly missing in the data source.

A mixed-design ANOVA was conducted to analyze changes in Python proficiency over time and differences across learner groups. The within-subjects factor was the specific assessment (1 to 30), and the between-subjects factor was learner group (Python experience, other programming experience, and no programming experience). This analysis assessed overall improvement across time points, differences in performance between groups, and interaction effects to determine whether any group demonstrated significantly greater or lesser gains. To further explore these effects, post-hoc pairwise t-tests with Bonferroni correction were conducted to examine contrasts between specific time points and group combinations. An additional ANCOVA was performed using Week 8 scores as the dependent variable, learner group as a fixed factor, and self-reported years of Python experience as a covariate, to assess whether group differences remained significant after accounting for prior experience. All statistical analyses were conducted using Python libraries including *pingouin*, *pandas*, and *seaborn*.

# 4. FIGURES/CAPTIONS

To assess the impact of accelerated Python instruction across learners with differing programming backgrounds, a mixed-design ANOVA was conducted across the 30 assessments as the within-subjects factor and GROUP (Have Programming Experience, No Python but Other Programming Experience, No Programming Experience) as the between-subjects factor. Weekly scores were scaled from 0–20 for comparability across timepoints

## 4.1 Main Effects and Interactions

Several statistically significant group differences emerged during the first few weeks of the course, prior to the completion of the Python instruction module, which concluded by the ninth assessment. These early assessments revealed performance gaps between learners based on their prior programming experience. For example, on the practice quiz on conditional statements, there was a significant main effect of group, $F(2, 114) = 4.05$, $p = .020$, with learners who had no programming experience scoring significantly lower than those with non-Python programming experience ($p = .018$). A similar pattern was observed on the practice quiz on looping statements, where group differences were again significant, $F(2, 105) = 4.90$, $p = .009$. In this case, learners with no programming experience scored significantly lower than both those with Python experience ($p = .009$) and those with other programming experience ($p = .034$).

This trend continued with the practice quiz on functions, where a significant group effect was found, $F(2, 105) = 3.63$, $p = .030$, and learners with no programming experience again performed worse than those with non-Python programming experience ($p = .034$). On the code evaluation assessment covering Python fundamentals, the group effect remained significant, $F(2, 98) = 3.28$, $p = .042$, with learners lacking any programming background scoring lower than those with Python experience ($p = .034$). A notable but unexpected finding appeared on the practice quiz covering the overview of generative AI, where a significant group difference emerged, $F(2, 164) = 4.18$, $p = .017$. Interestingly, learners with non-Python programming experience slightly outperformed those with Python experience ($p = .047$), although the magnitude of the difference was relatively small. There was no observed statistical difference between experienced Python programmers and those with no prior programming experience.

A statistically significant group difference also appeared on a practice quiz on large language models administered during Week six, $F(2, 116) = 3.97$, $p = .021$, with learners who had no programming experience scoring significantly lower than those with Python experience ($p = .017$). Notably, this assessment occurred two weeks after the completion of the Python instruction module and represents the only statistically significant group difference observed across the 21 assessments administered post-intervention. This isolated effect may reflect a lingering performance gap immediately following the transition to advanced content but does not indicate a sustained disadvantage for learners without prior programming experience.

Taken together, these findings indicate that the accelerated Python instruction—delivered alongside ChatGPT as a generative AI learning tool—was effective in rapidly upskilling learners with no prior programming experience. Early in the course, there was a clear, front-loaded advantage for learners with existing coding backgrounds, but this benefit diminished over time. Unlike the previously observed U-shaped learning curve, the updated analysis reveals a more stable or plateaued performance trajectory across groups, with significant differences attenuating by the midpoint of the course. Following the completion of the foundational Python module, learners from all backgrounds demonstrated comparable performance on subsequent assessments, reflecting successful adaptation and learning across the cohort. Notably, there were no consistent or sustained GROUP × TIME interactions, suggesting that all learners improved or stabilized at similar rates, regardless of prior experience. Only one statistically significant difference emerged across the 21 post-intervention assessments—a single practice quiz on large language models administered two weeks after the Python module concluded—further supporting the durability of learning gains. These results underscore the potential of AI-assisted instruction to close skill gaps quickly and equitably, enabling novices to catch up to their more experienced peers, even in technically demanding domains such as programming. This pattern is consistent with recent research suggesting that generative AI can serve as an effective tool for democratizing access to coding education [11, 12, 13, 17].

## 4.2 Post-hoc Comparisons

Post-hoc comparisons across the 30 assessment points provided insight into learners' progression over time. Rather than exhibiting sudden gains at specific intervals, learners demonstrated a gradual and sustained improvement in performance, particularly during the period of Python instruction. Effect sizes during this instructional window ranged from small to moderate ($\eta^2 = .06–.17$), indicating meaningful progress attributable to continued engagement with course content and project-based learning. Following the conclusion of the instructional module, learner performance stabilized at a higher level and remained consistent across subsequent assessments, suggesting effective skill acquisition and retention.

Across all assessments, no significant differences in overall performance were observed between the three learner groups (all $p > .10$), and effect sizes for group differences were consistently small ($\eta^2 < .05$), indicating that prior experience with Python or other programming languages did not yield a sustained advantage. While modest performance gaps were evident early in the course—particularly for learners without prior programming experience—these differences diminished over time. Only one statistically significant group-level difference emerged after the intervention, and even this isolated finding was associated with a small-to-moderate effect size. Taken together, these results reinforce the conclusion that the course structure, including the integration of generative AI tools such as ChatGPT, was effective in closing initial skill gaps and promoting equitable learning outcomes across diverse learner backgrounds

## 4.3 Qualitative Observations

In addition to quantitative performance metrics, feedback from industry mentors offers valuable qualitative insight into learners' experiences with Python instruction. Across multiple mentor reports, a common theme emerged: most learners appeared comfortable with Python by the end of the instructional module, even those with minimal prior experience. Mentors observed that learners frequently leveraged ChatGPT as a form of just-in-time support, using the tool to troubleshoot errors and reinforce concepts independently. This pattern of behavior aligns with prior research suggesting that generative AI can serve as an effective supplement to traditional programming instruction by delivering immediate, contextualized feedback [13, 14].

However, mentors also identified areas for instructional refinement. One mentor noted that learners with low programming experience struggled to engage with machine learning syntax and concepts, even when Python itself seemed manageable. This highlights a need to introduce foundational data literacy concepts—such as basic statistical visualizations and exploratory data analysis—earlier in the curriculum. Another mentor recommended decomposing Python code more deliberately during instruction, breaking down individual components to promote deeper understanding among learners without STEM backgrounds. This recommendation is particularly relevant for learners new to logic-based reasoning, who may benefit from a stepwise explanation of control flow, data structures, and function calls.

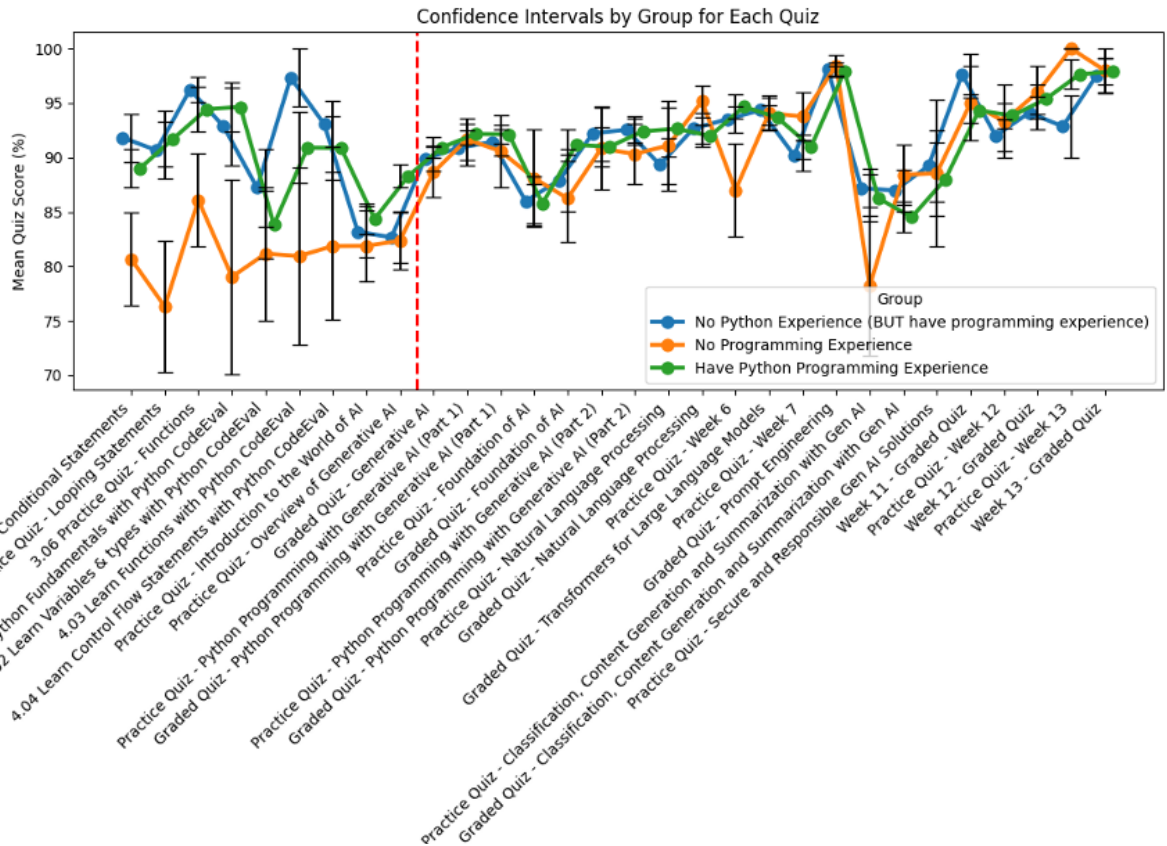

**Figure 1. Average Python proficiency score with 95% confidence interval over time.**

Several mentors suggested that Python examples in the initial MLS (Machine Learning Studio) sessions could be simplified to avoid introducing advanced constructs prematurely. Overly complex examples risk disengaging novices and may increase cognitive load during the early stages of learning. The same mentors proposed scaffolding more intentionally—such as incorporating formative quizzes, lightweight slide-based summaries, and opportunities for learner reflection—to reinforce key concepts and check comprehension before advancing.

Interestingly, one mentor noted that learners found it challenging to simultaneously learn Python and generative AI prompting during Week 2. This suggests a potential benefit to reordering the curriculum so that students develop confidence in AI prompting techniques before applying them in programming contexts. Such a revision could reduce early frustration and enhance the integration of metacognitive skills like question formulation and iterative refinement—both of which are essential for productive interaction with generative tools like ChatGPT.

Collectively, mentor feedback supports the broader conclusion of this study: generative AI can meaningfully reduce the entry barrier to programming, but the instructional environment must still be deliberately designed to address known learner challenges. By combining AI-assisted learning with structured scaffolding and responsive pedagogy, future iterations of the course may further enhance outcomes for learners across a range of backgrounds.

## 5. DISCUSSION

The findings from this study provide early evidence that generative AI tools, specifically ChatGPT, can support accelerated acquisition of foundational programming skills across learners with diverse backgrounds. Despite the highly compressed nature of the course—a two-week, asynchronous, self-paced Python module embedded within a sixteen-week generative AI curriculum—participants demonstrated measurable gains in programming proficiency. Notably, improvement was observed regardless of prior programming experience, underscoring the accessibility and scalability of the instructional model.

Interestingly, there were no significant differences in performance between learner groups over time. Learners with no prior programming experience performed comparably to those with backgrounds in Python or other languages. This finding suggests that prior exposure may not be a prerequisite for success in AI-supported instructional environments, particularly when learners are guided to use tools like ChatGPT to scaffold their understanding. The absence of a significant GROUP × WEEK interaction supports the conclusion that all groups improved at similar rates, and that the instructional design was broadly effective across experience levels. While a significant group-level difference occurred during the LLM multiple choice quiz between Python-experienced learners and programming novices ($p = .021$) may indicate a residual performance gap, the result did not continue across subsequent assessments and in fact, this group marginally outperformed those with prior programming experience during the practice quiz on RAG ($p = 0.091$).

These quantitative outcomes are echoed by qualitative feedback from industry mentors, who consistently reported that learners appeared comfortable with Python by the end of the instructional module. Several mentors observed that learners were increasingly adept at using ChatGPT to seek help, troubleshoot issues, and understand code, validating the model's utility as an on-demand instructional aid. However, mentors also identified common struggles—particularly in the early weeks—with managing the cognitive load of learning Python and prompting strategies simultaneously. One mentor suggested restructuring the course to teach prompting skills earlier, allowing learners to develop fluency with AI-assisted workflows before engaging in programming tasks.

The statistically significant improvement post-Python instruction intervention, particularly in the absence of continued formal Python instruction, highlights the potential for sustained skill development through self-guided practice and project-based learning. Mentor observations further support this, noting that learners were often capable of independently resolving syntax and logic issues through iterative engagement with ChatGPT. This aligns with prior work suggesting that AI-augmented learning environments can improve student outcomes while encouraging metacognitive growth [13, 14].

From a pedagogical perspective, these findings support the use of ChatGPT as a personalized learning assistant capable of addressing common barriers in programming education. As discussed in the background, novice programmers frequently struggle with syntax, semantics, debugging, and abstract thinking [1, 4, 7]. ChatGPT's ability to provide immediate natural language feedback, clarify code functionality, and assist with error handling offers a practical solution to these challenges—particularly in asynchronous, self-paced formats where instructor support is limited. By enabling learners to formulate questions, receive explanations, and iteratively refine their code, the model promotes not only programming competence but also metacognitive skills essential for lifelong learning.

At the same time, mentors emphasized the importance of instructional scaffolding. Several mentors reported that early examples used in coding exercises could be simplified to avoid overwhelming novices with advanced constructs too soon. Others recommended that mentors adopt a more granular, step-by-step explanation of each code block, especially for learners without a background in programming, math, or statistics. These suggestions highlight the need for intentional sequencing and clarity in instructional materials, particularly in AI-supported environments where learners may diverge in how they engage with feedback.

Additionally, mentors expressed that learners benefited from active questioning and narrative engagement from their instructors. Providing context through real-world anecdotes or interview-style discussions helped to sustain learner motivation and deepen understanding. These qualitative findings suggest that while ChatGPT can effectively automate many instructional tasks, human mentorship remains a critical element of learner success—especially in interpreting complex concepts and maintaining engagement.

While ChatGPT helped accelerate learning, it may also obscure opportunities for learners to build deeper debugging skills or fully internalize problem-solving strategies. Several mentors suggested incorporating lightweight assessments (e.g., quizzes) and visual aids (e.g., slides) to reinforce learning and promote retention. This reflects a broader pedagogical concern: overreliance on AI-generated solutions may hinder the development of independent reasoning, particularly if learners do not engage in reflection or deliberate practice.

It is also important to consider the generalizability of the findings. The study population consisted of highly motivated adult learners enrolled in a professional training program, many of whom were pursuing career advancement or technical specialization. While this context reflects the growing demand for workforce reskilling, results may differ in traditional undergraduate settings or among learners with lower intrinsic motivation. Furthermore, the assessments used in this study focused on fundamental Python knowledge and may not fully capture deeper competencies such as algorithmic thinking or code optimization.

Future research should examine long-term retention and transfer of skills, particularly in settings where learners are expected to independently design and implement software solutions. Additionally, studies comparing AI-augmented instruction with traditional or hybrid models would offer valuable insight into tradeoffs in learning outcomes, student engagement, and instructional efficiency. Experimental designs incorporating behavioral logging, eye tracking, or think-aloud protocols could further elucidate how learners interact with generative tools and when such tools are most effective.

In conclusion, the findings of this study suggest that generative AI can play a meaningful role in accelerating programming instruction, reducing barriers to entry, and supporting scalable, inclusive models of technical education. When integrated thoughtfully, tools like ChatGPT can help bridge gaps in prior experience and empower learners to develop essential digital competencies. As educational institutions and employers continue to seek agile, high-impact training solutions, AI-augmented instruction holds promise as a key strategy for meeting the demands of the digital economy